\documentclass[aps,prl,twocolumn,superscriptaddress,showpacs,preprintnumbers,amsmath,amssymb]{revtex4}
\usepackage{graphicx}
\usepackage{dcolumn}
\usepackage{bm}
\usepackage{multirow}
\usepackage{amsbsy}
\usepackage{subfigure}
\newcommand\T{\rule{0pt} {2.6ex}}
\newcommand\B{\rule[-1.2ex] {0pt} {0pt} }

\begin{document}


\title{Coupled magnetic and ferroelectric domains in multiferroic $\mathbf{Ni_3V_2O_8}$}

\author{I. Cabrera}
\affiliation{Department of Physics and Astronomy, Johns Hopkins University, Baltimore, Maryland 21218, USA}
\affiliation{National Institute of Standards and Technology, Gaithersburg, Maryland 20899, USA}
\author{M. Kenzelmann}
\affiliation{Laboratory for Developments and Methods, Paul Scherrer Institute, CH-5232 Villigen, Switzerland}
\author{G. Lawes}
\affiliation{Department of Physics and Astronomy, Wayne State University, Detroit, Michigan 48201, USA}
\author{Y. Chen}
\author{W. C. Chen}
\author{R.~Erwin}
\affiliation{National Institute of Standards and Technology, Gaithersburg, Maryland 20899, USA}
\author{T.~R.~Gentile}
\affiliation{National Institute of Standards and Technology, Gaithersburg, Maryland 20899, USA}
\author{J. B. Le\~ao}
\affiliation{National Institute of Standards and Technology, Gaithersburg, Maryland 20899, USA}
\author{J. W. Lynn}
\affiliation{National Institute of Standards and Technology, Gaithersburg, Maryland 20899, USA}
\author{N. Rogado}
\affiliation{DuPont Central Research and Development, Experimental Station, Wilmington, Delaware 19880, USA}
\author{R. J. Cava}
\affiliation{ Department of Chemistry and Princeton Materials Institute, Princeton University, Princeton, New Jersey 08544, USA}
\author{C. Broholm}
\affiliation{Department of Physics and Astronomy, Johns Hopkins University, Baltimore, Maryland 21218, USA}
\affiliation{National Institute of Standards and Technology, Gaithersburg, Maryland 20899, USA}

\date{\today}

\begin{abstract}

Electric control of multiferroic domains is demonstrated through polarized magnetic neutron diffraction. Cooling to the cycloidal multiferroic phase of Ni$_3$V$_2$O$_8$ in an electric field $\mathbf{E}$ causes the incommensurate Bragg reflections to become neutron spin polarizing, the sense of neutron polarization reversing with $\mathbf{E}$. Quantitative analysis indicates the $\mathbf{E}$-treated sample has handedness that can be reversed by  $\mathbf{E}$.  We further show close association between cycloidal and ferroelectric domains through $\mathbf{E}$-driven spin and electric polarization hysteresis.  We suggest that definite cycloidal handedness is achieved through magneto-elastically induced Dzyaloshinskii-Moriya interactions.
\end{abstract}


\maketitle
 
Materials that are both ferroelectric and magnetic are classified as multiferroics.  In some multiferroics where the ferroelectric and magnetic phases coexist, spin and charge are strongly coupled, leading to the possibility of controlling magnetic properties through an electric field $\mathbf{E}$.  Such a nonlinear magneto-electric response is of fundamental interest and holds the potential for applications that include sensing, spintronics, and microwave communication \cite{Sousa2008Multiferroic-Ma}.  Recent studies have shown that an external $\mathbf{E}$ applied to multiferroics with non-collinear spin structures, such as TbMnO$_3$ and LiCu$_2$O$_2$, favors a particular handedness of the magnetic order \cite{yamasaki:147204,seki:127201}.  Other studies have shown the $\mathbf{E}$ control of domain population related to equivalent magnetic propagation vectors \cite{lebeugle:227602}.  Here we examine the suppression and promotion of cycloidal magnetic structures in Ni$_3$V$_2$O$_8$ (NVO) by an applied $\mathbf{E}$.  Our quantitative analysis of the polarized magnetic diffraction cross-section and hysteresis curve for this multiferroic material indicates that a clockwise cycloidal single crystal can be generated and is stabilized by magneto-elastically induced Dzyaloshinskii-Moriya interactions. 

\begin{figure}
\includegraphics[width=\linewidth,clip=true,trim= 0 0 0 0]{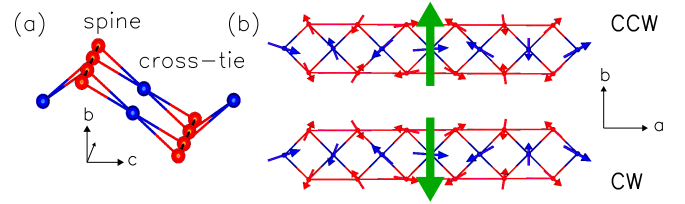} 
\caption{\label{NVOstructure:epsart} (Color online) (a) NVO crystal sublattice showing Ni$^{2+}$ spine (red) and cross-tie (blue) sites.  (b) Counter-clockwise (top) and clockwise (bottom) spin cycloids propagating along the $\mathbf{a}$ axis.  The (green) vertical arrow indicates the direction of $\mathbf{P}$.}
\end{figure}

NVO is an insulating magnet with spin-1 Ni$^{2+}$ ions arranged in a buckled kagom\'e-staircase geometry \cite{lawes:087205}.  The spins occupy two distinct crystallographic sites denoted cross-tie and spine [See Fig. \ref{NVOstructure:epsart} (a)].  Competing nearest and next-nearest neighbor interactions along the spines yield a complex magnetic phase diagram \cite{Lawes2004Competing-Magne}.  Magnetic inversion symmetry breaking was inferred in the so-called low-temperature incommensurate (LTI) phase, where unpolarized neutron diffraction data indicate a magnetic cycloidal structure with spins in the \mbox{\textbf{a}-\textbf{b}} plane and pyrocurrent measurements find concomitant electric polarization along the \textbf{b} axis.  A Landau mean field theory was previously devised to account for this multiferroic behavior \cite{lawes:087205,kenzelmann:014429,harris-2006}.  The free-energy expansion is
\begin{equation}
\begin{split}
F=a(T-T_H)\sigma^2_H + b(T-T_L)\sigma^2_L + \mathcal{O}(\sigma^4)\\ + (2\chi_E)^{-1}\mathbf{P}^2 + V.
\end{split}
\end{equation}
Here, $a$ and $b$ are constants, $T$ is temperature, $\sigma_H$ and $\sigma_L$ are the magnetic order parameters in the high-$T$ incommensurate and LTI phases, respectively, 
$\chi_E$ is the electric susceptibility, and $\mathbf{P}$ is the electric polarization.  The last term is the lowest order (trilinear) symmetry-allowed multiferroic interaction, which in the LTI phase is given by $V_{LTI}=\sum_{\substack{\gamma}}a_\gamma\sigma_H\sigma_LP_\gamma$.  Minimizing $F$ with respect to $\mathbf{P}$, one finds that $P_b=\bf\hat{b}\cdot \bf P$ varies with $T$ in proportion to the product of the two magnetic order parameters $(P_{b} \propto a_{b} \chi_{E} \sigma_{L} \sigma_{H})$, as observed experimentally \cite{Lawes2008Magnetically-in}.  The theory also suggests that ferroelectric and magnetic domains are coupled in NVO.  Here we examine this hypothesis by probing the magnetic and ferroelectric response to an applied $\mathbf{E}$ in the multiferroic phase.

NVO crystals were grown from a BaO-V$_2$O$_5$ flux \cite{Lawes2004Competing-Magne}.  The buckled kagom\'e layers span the \mbox{\textbf{a}-\textbf{c}} crystallographic plane and form the largest crystalline surfaces.  A 0.58 g, 120 mm$^3$ crystal was selected for this experiment.  A parallel-plate capacitor was formed by evaporating a 5 nm Cr/40 nm Au layer on each large face of the crystal, hence normal to the ferroelectric axis.  Au wires were attached to each side of the sample using silver epoxy paste.  Polarized neutron diffraction measurements were carried out on BT-7 at NIST.  A 14.7 meV neutron beam was polarized and analyzed by $^3$He neutron spin filters \cite{WCChen-07}.  Helmholtz coils were used to generate a guide field at the sample position, thus defining the neutron spin quantization axis.  In the vertical field (VF) configuration the field strength was 0.4 mT, oriented normal to the scattering plane to within 0.07 rad.  In the horizontal field (HF) configuration, the field strength was 0.8 mT, oriented within 0.1 rad of the horizontal plane and parallel to wave vector transfer $\mathbf{Q=k_i-k_f}$ to within 0.2 rad.  Mezei neutron spin-flippers were mounted before and after the sample, providing access to a total of eight configurations for the incoming and outgoing neutron spin.  The nomenclature used is as follows: (+) refers to a flipper off and ($-$) refers to a flipper on.  With both flippers off the neutron spin nominally points up for VF and parallel to $\mathbf{Q}$ for the HF configuration.  
\begin{figure}
\includegraphics[width=\linewidth,clip=true, trim= 0 0 0 0]{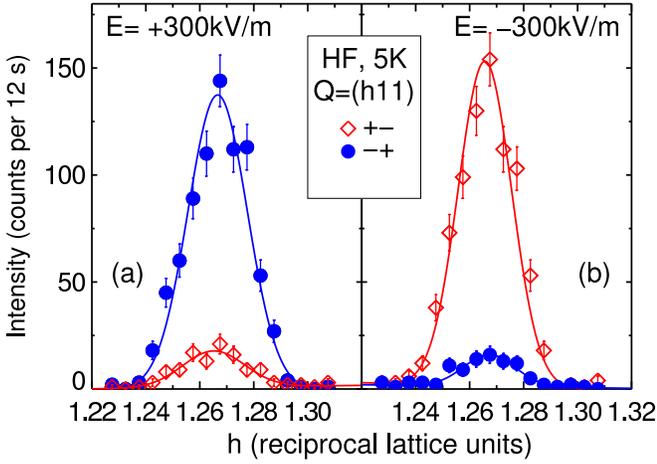} 
\caption{\label{all300VLegendCorrected:epsart}(Color online) Full polarized magnetic diffraction in an electric field under opposite spin-flip scattering conditions.  Statistical uncertainties represent one standard deviation.  (a) Cooling in a field $\mathbf{E}$= +300 kV/m from 11 K to 5 K ($\mathbf{E}$-cooling) favors cycloidal domains that predominantly diffract neutrons polarized antiparallel to $\mathbf{Q}$. (b) Reversing $\mathbf{E}$ yields the opposite polarized intensity asymmetry.}
 \end{figure}
 
HF spin-polarized diffraction data are shown in Fig.\ \ref{all300VLegendCorrected:epsart}.  There is a strong asymmetry in the intensity between the \mbox{(+/$-$)} and \mbox{($-$/+)} spin-flipper configurations which reverses with $\mathbf{E}$.  This demonstrates in a qualitative fashion the \textbf{E}-driven suppression and promotion of cycloidal magnetic domains \cite{Blume1963}.  Examining the data more carefully, we note that a finite peak remains under the (+/$-$)[($-$/+)] spin-flip scattering condition, even after cooling in a +300 [$-$300] kV/m field.  Similar effects have been seen in \cite{yamasaki:147204,seki:127201}, but have not been fully accounted for.  Note that, because the strong diffraction cross-section is suppressed by both the incident and the final beam $^3$He neutron spin filters, the residual intensity cannot be accounted for by finite beam polarization.  Quantitative analysis of the polarized diffraction cross-section is needed to account for this effect. 

Within a single domain, the magnetic structure of NVO in the LTI phase can be described as follows:
\begin{equation}
\label{spinstructure}
\mathbf{S_{Rd}}_i=\mathbf{m_{d}}_i e^{i\mathbf{q_m}\cdot(\mathbf{R+{d}}_i)}+\mathbf{m_{d}}^*_ie^{-i \mathbf{q_m}\cdot(\mathbf{R+{d}}_i)}.
\end{equation}
Here, $\mathbf{q_m}$ is the magnetic propagation vector, $\mathbf{R}$ is a vector from the origin to the unit cell, $\mathbf{d}_i$ are position vectors for Ni$^{2+}$ ions within the conventional unit cell, and $\mathbf{m_{d}}_i$ transform according to irreducible representations of the magnetic space group and specify the time-averaged magnetic moments on Ni$^{2+}$ sites.  In the LTI phase, where electric polarization is present, the spin structure was previously described by the $\Gamma_1$ and $\Gamma_4$ irreducible representations with best fit basis vectors for spine and cross-tie sites $\mathbf{m_{s}}_i$ and $\mathbf{m_{c}}_i$, $i=1,4$ \cite{kenzelmann:014429}.  The resulting spin structure is a clockwise (CW) cycloid, progressing along $\mathbf{a}$ [see Fig.\ \ref{NVOstructure:epsart} (b)].  Spatial inversion is a symmetry operation of the paramagnetic phase that converts a CW cycloid into a counterclockwise (CCW) cycloid.  The set of complex $\mathbf{m_{d}}_i$ for CW and CCW cycloids are listed in Table \ref{spincomp}.  We expect that domains in NVO are associated with these symmetry-related structures.
\begin{table}
\footnotesize
\caption{\label{spincomp} Spin components on Ni$^{2+}$ spine (s) and cross-tie (c) sites describing CW and CCW cycloids.  The inversion operator $\mathcal{I}$ converts a CW cycloid into a CCW as follows: for  $\alpha=a$ or $c$, $\mathcal{I}[m_{si}^{\alpha}]=\mp(m_{si}^{\alpha})^*$ and for the b-component, $\mathcal{I}[m_{si}^{b}]=\pm(m_{si}^{b})^*$ (upper sign for $i=\Gamma_1$ and lower sign for $i=\Gamma_4$).  For all cross-tie sites $\mathcal{I}[m_{ci}]=(m_{ci})^*$. There are six additional atoms in the conventional unit cell, obtained by translating $d_i$ by ($\frac{1}{2},\frac{1}{2},0$). Asterisks denote complex conjugation.} 
\begin{ruledtabular}
\begin{tabular}{c  c  l}
$d_i$ & $\mathbf{m_{d}}_i$&$\mathbf{m_{d}}_i$\\ 
$=(l, m, n)$ &(CW) & (CCW)\\[1ex]
\hline 
($\frac{1}{4}$, -0.13, $\frac{1}{4}$)\T &$\mathbf{m_{d_1}}=(m_{s1}^a,m_{s1}^b,m_{s1}^c)+(m_{s4}^a,m_{s4}^b,m_{s4}^c)$&$(\mathbf{m_{d_3}})^*$\\ [0.7ex]
($\frac{1}{4}$, 0.13, $\frac{3}{4}$)&$\mathbf{m_{d_2}}=(m_{s1}^a,\bar m_{s1}^b,\bar m_{s1}^c)+(\bar m_{s4}^a,m_{s4}^b,m_{s4}^c)$&$(\mathbf{m_{d_4}})^*$\\ [0.7ex]
($\frac{3}{4}$, 0.13, $\frac{3}{4}$)&$\mathbf{m_{d_3}}=(\bar m_{s1}^a,m_{s1}^b,\bar m_{s1}^c)+(m_{s4}^a,\bar m_{s4}^b,m_{s4}^c)$&$(\mathbf{m_{d_1}})^*$\\[0.7ex]
($\frac{3}{4}$, -0.13, $\frac{1}{4}$)&$\mathbf{m_{d_4}}=(\bar m_{s1}^a,\bar m_{s1}^b,m_{s1}^c)+(\bar m_{s4}^a,\bar m_{s4}^b,m_{s4}^c)$&$(\mathbf{m_{d_2}})^*$\\[0.7ex]
(0, 0, 0)&$\mathbf{m_{d_5}}=(m_{c1}^a,0,0)+(0,m_{c4}^b,m_{c4}^c)$&$(\mathbf{m_{d_5}})^*$\\[0.7ex]
($\frac{1}{2}$, 0, $\frac{1}{2}$)\B&$\mathbf{m_{d_6}}=(\bar m_{c1}^a,0,0)+(0,\bar m_{c4}^b,m_{c4}^c)$&$(\mathbf{m_{d_6}})^*$
\end{tabular}
\end{ruledtabular}
\end{table}

\begin{figure}
\includegraphics[width=\linewidth,clip=true,trim= 0 10 0 0]{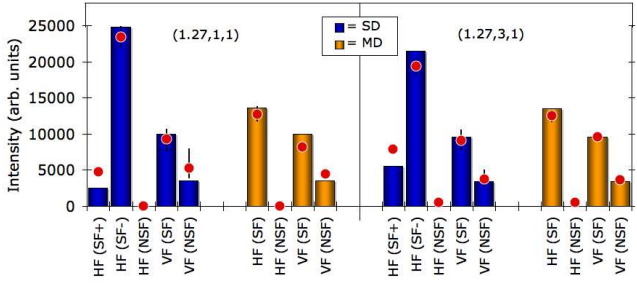}  
\caption{\label{ComparedIntensities:epsart} (Color online) Comparison between measured and calculated intensities for (1.27,1,1) and (1.27,3,1) peaks.  Columns represent calculated intensities for single-domain (SD) and multi-domain (MD) cycloidal structures.  (Red) circles denote the average measured intensity and vertical bars show lowest and highest values measured for each particular condition.  The categories include spin-flip (SF) and non-spin-flip (NSF) modes under HF and VF.  + and $-$ refer to the sign of $\mathbf{P_i}\cdot\mathbf{E}$.  Polarization efficiency corrections were applied in this comparison \cite{chuck91}.}
\end{figure}

To quantitatively characterize the cycloidal domains, we employed Blume's equations for elastic scattering of polarized neutrons \cite{Blume1963}.  The polarization-resolved differential diffraction cross-section is:
\begin{multline}
\label{spinpol}
\frac{d\sigma}{d\Omega}\mathbf{P_f}= (\gamma r_0)^2\frac{N(2\pi)^3}{v_0}|F_{\mathrm{Ni}}(\boldsymbol{\tau})|^2
\{[\mathbf{P_i}\cdot\mathbf{\mathcal{F}_\perp^*(\boldsymbol{\tau})}] \mathbf{\mathcal{F}_\perp}(\boldsymbol{\tau})\\
+\mathbf{\mathcal{F}_\perp^*(\boldsymbol{\tau})}[\mathbf{P_i}\cdot\mathbf{\mathcal{F}_\perp(\boldsymbol{\tau})}]
-\mathbf{P_i}|\mathbf{\mathcal{F}_\perp(\boldsymbol{\tau})}|^2\\
\mp i[\mathbf{\mathcal{F}_\perp^*(\boldsymbol{\tau})}\times\mathbf{\mathcal{F}_\perp(\boldsymbol{\tau})}]\}  
\delta(\mathbf{Q}\pm\mathbf{q_m}-\boldsymbol{\tau}).
\end{multline}
Here, $\gamma$ =1.913, $r_0$ is the classical electron radius,  $N$ is the number of magnetic unit cells with volume $v_0$, $F_{\mathrm{Ni}}(\boldsymbol{\tau})$ is the form factor for Ni$^{2+}$,  $\boldsymbol{\tau}$ is the nuclear reciprocal lattice vector hosting the magnetic satellite, $\mathbf{P_i}$ and $\mathbf{P_f}$ are the initial and final neutron spin polarization, respectively,  $\mathbf{\mathcal{F}_\perp(\boldsymbol{\tau)}}=\hat{\mathbf{Q}}\times(\mathbf{\mathcal{F}(\boldsymbol{\tau})}\times\hat{\mathbf{Q}})$, and the magnetic vector structure factor is given by
 \begin{equation}
 \mathbf{\mathcal{F}}(\boldsymbol{\tau})=\sum_{d_i}\frac{\mathbf{m_{d}}_i}{\mu_B} \exp[{i\boldsymbol{\tau}\cdot \mathbf{d}_i}]. \end{equation}
Furthermore, the overall $\mathbf{P_i}$-dependent diffraction cross-section is:
 \begin{multline}
 \label{diffcrosssect}
 \frac{d\sigma}{d\Omega}= (\gamma r_0)^2\frac{N(2\pi)^3}{v_0}|F_{\mathrm{Ni}}(\boldsymbol{\tau})|^2\{|\mathcal{F_\perp(\boldsymbol{\tau)}}|^2\\
 \pm i\mathbf{P_i}\cdot[\mathcal{F_\perp^*(\boldsymbol{\tau)}}\times\mathcal{F_\perp(\boldsymbol{\tau)}}]\}\delta(\mathbf{Q}\pm\mathbf{q_m}-\boldsymbol{\tau}).
\end{multline} 
Dividing Eq.\ (\ref{spinpol}) by (\ref{diffcrosssect}) provides the final polarization unit vector $\mathbf{P_f}$.  The calculated intensities for a single-domain (SD) cycloidal structure (CW or CCW) are readily obtained using Table \ref{spincomp}.  A multi-domain (MD) cycloidal structure is described by averaging the calculated intensities for the CW and CCW spin structures.  These results are compared to the measured polarized beam intensities in Fig.\ \ref{ComparedIntensities:epsart} for different domain population (SD and MD), field conditions (HF and VF), and spin-flipper configurations.  Measured intensities in the absence of $\mathbf{E}$ agree with the calculated MD values, indicating equal CW and CCW volume fractions.  The polarized beam data are in excellent agreement with the intensities calculated on the basis of the previously published structure \cite{kenzelmann:014429}.  The analysis accounts for the weak peaks in Fig. \ref {all300VLegendCorrected:epsart} and allows us to conclude that $\mathbf{E}$ applied along $\pm\mathbf{b}$ induces a CCW(CW) cycloid, as defined in Fig. \ref{NVOstructure:epsart} (b).  
  
\begin{figure}
 \includegraphics[width=\linewidth,clip=true, trim= 0 0 0 0]{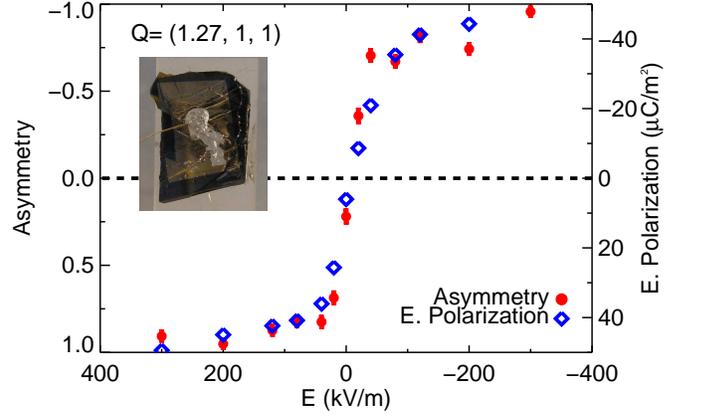}  \caption{\label{EpolAsymVsE:epsart} (Color online) Evolution of magnetic and ferroelectric domains after $\mathbf{E}$-cooling the sample.  Ferroelectric data was obtained by integrating the pyroelectric current while $\mathbf{E}$-heating to the paramagnetic phase at a rate of 1 K/min.  (Insert) Gold-plated NVO single crystal.}
\end{figure}

We now examine the evolution of ferroelectric and cycloidal (i.e., multiferroic) domains with different electric field protocols, and show that these multiferroic domains in NVO have a defined magnetic and electric state that is directly coupled.  Fig.\ \ref{EpolAsymVsE:epsart} shows polarized diffraction and pyroelectric current data sensitive to the volume fraction of magnetic and ferroelectric domains, respectively, after $\mathbf{E}$-cooling the sample.  The magnetic diffraction data are presented in terms of the CW-CCW cycloid asymmetry ($A$):
\begin{equation}
A=\frac{(I_{\mp}) - (I_{\pm})}{(I_{\pm}) + (I_{\mp})}\cdot\frac{(\sigma^{\pm}_{CW}) + (\sigma^{\pm}_{CCW})}{(\sigma^{\pm}_{CW}) - (\sigma^{\pm}_{CCW})}.
\end{equation}
Here $I_{\mp}(I_{\pm})$ is the magnetic Bragg peak intensity under the on/off (off/on) spin-flipper configurations and $\sigma^{\pm}_{CW}(\sigma^{\pm}_{CCW})$ is the calculated cross-section for CW (CCW) cycloids in the off/on configuration.  This expression correctly accounts for the contribution of each cycloid to  $I_{\mp}$ and $I_{\pm}$ .  The data show that by cooling in a sufficiently high $\mathbf{E}$ oriented along the $\mathbf{-b}$ direction, 98 \% of the long-range ordered part of the sample contains CW cycloids (the direction of $\mathbf{E}$ was confirmed in a separate experiment).  The maximum spontaneous electric polarization attained was \mbox{$\sim$ 47 $\mu$C/m$^2$}.  This is 46 \%  lower than previously observed for NVO \cite{lawes:087205}.  The discrepancy may reflect uncertainty in determining the surface area for the small single crystalline sample or sample variability.   

\begin{figure}
 \includegraphics[width=\linewidth,clip=true, trim= 0 0 0 0]{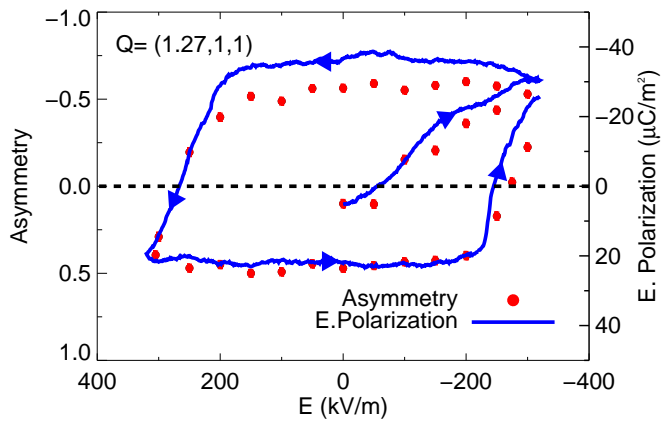} 
 \caption{\label{EpolAsymHyst320kVoverm:epsart} (Color online) Hysteresis loop for magnetic asymmetry and electric polarization versus $\mathbf{E}$.  The sample was zero-field cooled to 5.0(2) K.  Hysteresis data were obtained while varying $\mathbf{E}$ as indicated by the arrows.  Typical waiting time before neutron diffraction measurements was 10 s and average ramp rate was 2 kV/(m$\cdot$s).  Ferroelectric hysteresis data were obtained with the sample in an evacuated can (to avoid electric breakdown in helium exchange gas) by integrating the voltage-step induced current while varying $\mathbf{E}$ in steps of 1 kV/m at an average rate of 0.11 kV/(m$\cdot$s).}
 \end{figure}

Fig.\ \ref{EpolAsymHyst320kVoverm:epsart} shows the $\mathbf{E}$-dependence of the magnetic and ferroelectric volume fractions after zero-field cooling the sample into the LTI phase.  The product of coercive field and saturation polarization for NVO is $1.05\times 10^{-5}$ J$\cdot$cm$^{-3}$.  This value is similar to multiferroic MnWO$_4$ \cite{kundys:172402}, but more than 4 orders of magnitude smaller than for the bulk ferroelectric Pb(Zn-Ti)O$_3$ and the thin-film ferroelectric BiFeO$_3$ \cite{2007ApPhA..89..737Y,Dho2006}.  This reflects the different origin and much smaller magnitude of ferroelectric lattice distortion in inversion symmetry-breaking frustrated magnets.  Excellent correspondence between the $\mathbf{E}$-dependent electric and magnetic domain fractions under field-cooled and zero-field cooled conditions leads us to conclude that these domains coincide in NVO.  Furthermore, these results suggest that the magnetic and electric domain walls are coupled in this system.  This is different from multiferroic hexagonal YMnO$_3$, where magnetic and ferroelectric domain walls can move independently \cite{fiebig2002}.  NVO's  multiferroic domains should have interesting domain wall characteristics that could be imaged with magneto-optical techniques \cite{Fiebig:05}.

While the LTI cycloids could result from competing Heisenberg interactions along the $\mathbf{a}$ axis, definite spin handedness throughout the sample as demonstrated here can only originate from an interaction like the Dzyaloshinskii-Moriya (DM) of the form $\mathbf{D}\cdot(\mathbf{S}_n\times \mathbf{S}_{n+1})$ with $\mathbf{D}\cdot\mathbf{\hat a} \neq 0$, which has opposite expectation values for the CW and CCW cycloids.  The importance of DM interactions was previously pointed out for BiFeO$_3$ \cite{lebeugle:227602}.  Such a term is forbidden by symmetry in the paramagnetic phase of NVO \cite{sergienko:094434}, but apparently must be present in the Hamiltonian for the LTI phase.  We infer that magneto-elastic distortions in this phase generate DM interactions that select a definite cycloidal handedness. 

In summary, we have demonstrated electric control of cycloidal handedness in multiferroic NVO through polarized magnetic neutron diffraction.  The previously determined spin structure is consistent with the polarized diffraction data and our quantitative analysis allows us to conclude that a single domain cycloidal structure can be generated through $\mathbf{E}$-cooling to the level of 98 \%.  We observed comparable ferroelectric polarization and cycloidal magnetic hysteresis, indicating that electric and magnetic domains are coupled, as predicted by the trilinear theory.  The results suggest that domains with electric polarization along $\pm\mathbf{b}$ feature DM interactions that select cycloids with CCW(CW) handedness.  

\begin{acknowledgments}
We gratefully acknowledge discussions with O. Tchernyshyov.  Work at JHU was supported by NSF through Grant No. DMR-0706553.  G. L. acknowledges support from the NSF through Grant No. DMR-0644823.  M. K. was supported by Swiss NSF contract PP002-102831.  The development and application of 3He spin filters was supported in part by the U. S. Dept. of Energy, Basic Energy Sciences.
\end{acknowledgments}

\end{document}